\documentclass[preprint,showpacs,preprintnumbers,amsmath,amssymb]{revtex4}

\usepackage{graphicx}
\usepackage{dcolumn}
\usepackage{bm}

\begin{document}

\title{  Gravitational potential of a point mass in a brane world}

\author{Rom\'an Linares$^1$}
\email{lirr@xanum.uam.mx}
\author{Hugo A. Morales-T\'ecotl$^1$}
\email{hugo@xanum.uam.mx}
\author{Omar Pedraza$^2$}
\email{omarp@uaeh.edu.mx}
\author{Luis O. Pimentel$^1$}
\email{lopr@xanum.uam.mx}

\affiliation{$^1$ Departamento de F\'{\i}sica, Universidad Aut\'onoma Metropolitana Iztapalapa,\\
San Rafael Atlixco 186, C.P. 09340, M\'exico D.F., M\'exico,}

\affiliation{$^2$
\'Area Acad\'emica de Matem\'aticas y F\'isica, Universidad Aut\'onoma del Estado de Hidalgo, \\
Carretera Pachuca-Tulancingo Km. 4.5, C P. 42184, Pachuca,
M\'exico.}


\begin{abstract}
In brane world models, combining the extra dimensional field modes   with the standard four dimensional ones  
yields interesting physical consequences  that have been proved from high energy physics to cosmology. Even some 
low energy phenomena have been considered along these lines to set bounds on the brane model parameters.
In this work we extend to the gravitational realm a previous result which gave finite electromagnetic and 
scalar  potentials and self energies for a source looking pointlike to an observer sitting in a 4D Minkowski subspace 
of the single brane of a Randall-Sundrum spacetime including compact dimensions. We calculate here  
the gravitational field for the same type of source by solving the linearized Einstein equations. Remarkably, it turns 
out  to be also non singular. Moreover, we use gravitational experimental results of the Cavendish type and the 
Parameterized Post Newtonian  (PPN) coefficients, to look for admissible values of the brane model parameters. The 
anti de Sitter radius hereby obtained is concordant with previous results based on Lamb shift in hydrogen. However, 
the resulting PPN parameters lie outside the acceptable value domain.  
\end{abstract}


\maketitle
\section{Introduction}

We are close to celebrate 100 years of the birth of General Relativity (GR), one of the most beautiful and spectacular 
theories in physics ever conceived. GR is beautiful because at the time, it introduced deep and unexpected physical 
concepts that allowed to understand the gravitational field and its relation with the geometry of the space-time. It is 
spectacular because, despite its age, the GR equations of motion have remained immutable in form and they 
describe with  great accuracy most of the observable gravitational physics. Over the years GR has passed most of the 
experimental tests concerning the theory, however it is well known that there exist some phenomena escaping an 
accurate description within the framework of GR and the Standard Model of particle physics such as dark matter and 
dark energy. In order to develop a consistent theory that could describe these kind of phenomena, physicists have 
tried to modify either GR or Quantum Mechanics and consider possible extensions of the Standard Model. Indeed, 
the effort to place limits on possible deviations from the standard formulations of such theories continue nowadays. 

Since its inception there have been many attempts to modify GR with different purposes. Soon after its conception 
there were notable proposals with the idea to extend it and incorporate it in a larger unified theory. A relevant 
example for the purpose of this work is the higher dimensional theory introduced by Kaluza \cite{Kaluza:1921tu} and 
refined by Klein \cite{Klein:1926tv}. More recently and with the aim to solve the hierarchy problem, the ideas of large 
extra dimensions  \cite{Antoniadis:1998ig,Arkani-Hamed:1998rs,Antoniadis:1990ew} and brane worlds were 
introduced \cite{Randall:1999ee,Randall:1999vf,Visser:1985qm,Rubakov:1983bb,Akama:1982jy}. Of course in the 
literature there are many other attempts to modify GR (see e.g. \cite{Clifton:2011jh} and references therein) and 
currently people continue exploring the physical consequences predicted from them all and most important 
confronting them with experimental data. This work follows the same strategy, we will explore a particular 
characteristic of the gravitational field, specifically the behavior of the gravitational potential generated by a {\it point 
like} source in the so called RSII$p$ model which modifies GR by including extra dimensions, and we will confront it 
with experimental data available today.

The RSII$p$ model is an extension of the 5D Randall-Sundrum (RS) model with one brane (RSII) extended by 
$p$ compact extra dimensions (RSII$p$) \cite{Dubovsky:2000av,Oda:2000zc,Dubovsky:2000am}, its 
construction was motivated for the need to improve the localization properties of matter fields within the standard 
RS model. Specifically in the RSII model there exists a problem to localize spin 1 fields on the brane 
and a way out to this problem can be achieved by extending the model with $p$ compact dimensions  
\cite{Dubovsky:2001pe,Rubakov:2001kp}. Thus the RSII$p$ setup contains all the nice features of the RS model and 
additionally has the advantage to localize every kind of field on the brane. These higher dimensional models have 
the property to modify gravity in the low scale length regime and a huge amount of physical phenomena have been 
studied over the years, ranging from particle physics (see e.g. \cite{Allanach:2006fy,Rizzo:2010zf} and references 
therein) to cosmology (see e.g. \cite{Maartens:2010ar,Maartens:2003tw} and references therein). Moreover, recently it was shown that an electric source lying in the single brane of a 
RSII$p$ spacetime which looks pointlike to an observer sitting in usual 3D space, produces a static potential which is 
nonsingular at the 3D point position \cite{Linares:2011sg,MoralesTecotl:2006eh} and, furthermore, it matches Coulomb potential outside a small neighborhood. Amusingly, coping with classical singularities goes back to 
the non-linear proposal made by Born and Infeld \cite{Born:1934gh}. 
In regard to the divergences in field theory, over the years there have been many attempts to formulate 
a theory that avoids the problem, or at least that  could improve, for instance,  the high energy behavior of GR. Among them 
we have for instance: String Theory  (see e.g.  \cite{Polchinski:1998rq} and references therein), non commutative 
theories (see e.g. \cite{Szabo:2001kg} and references therein) and the recent attempt made by Horava 
\cite{Horava:2009uw} of a modified UV theory of gravity.

In this work we extend the analysis of  \cite{Linares:2011sg} to the case of the gravitational field. As we will show the 
classical potential due to an effective  4D punctual source becomes regular at the position of the source in analogous 
way to the scalar and gauge cases. To complement our study we compare the consequences of this feature with 
some experimental observations, in particular  we have chosen to compare the predictions of the model with the 
experimental data of a Cavendish type experiment which imposes a bound to the anti de Sitter (adS) radius of the 
bulk adS metric. We also obtain the Parameterized Post Newtonian (PPN) coefficients of the resulting effective 
theory. 

The paper is organized as follows. In section \ref{background} we describe briefly the RSII$p$ scenarios. In section 
\ref{GravitationalCase}  we discuss the linearized Einstein equations in the low energy regime for a massive particle 
with the topology of $T^p$ torus, but which is seen as punctual by an observer in our 4D world. In section 
\ref{solutions} we obtain the metric perturbations and in section \ref{exper} we discuss a Cavendish type experiment 
and we give the PPN coefficients. We give a short discussion of our results in section \ref{conclu}.

\section{Randall-Sundrum II$p$ scenarios}\label{background}

The way in which the Randall-Sundrum II$p$ (RSII$p$) scenarios arise has been discussed several times in the 
literature (see e.g. \cite{Dubovsky:2000av,Oda:2000zc,Dubovsky:2000am,Dubovsky:2001pe,Rubakov:2001kp}), 
so here we just give a short summary including the most important features of the model. The RSII$p$ setups consist 
of a ($3+p$)-brane with $p$ compact dimensions and positive tension $\sigma$, embedded in a ($5+p$) spacetime 
whose metrics are two patches of anti-de Sitter (AdS$_{5+p}$) having curvature radius $\epsilon$ (for convenience 
in some equations we will use instead of the radius, its inverse: $\kappa \equiv \epsilon^{-1} $). The 
model arise from considering the $(5+p)$D Einstein action with bulk cosmological constant $\Lambda$ and the 
action of a  ($3+p$)-brane
\begin{equation}\label{action}
S=\frac{1}{16 \pi G_{5+p}} \int d^4 x \, dy \prod_{i=1}^{p} R_i d \theta_i \sqrt{|g^{(5+p)}|}  \,  
\left( R^{(5+p)} - 2  \Lambda \right)  +  S_{brane} ,
\end{equation}
which leads to the Einstein equations of motion
\begin{equation}\label{EqsMotion}
R_{MN}-\frac{1}{2} R \, g^{(5+p)}_{MN} +  \Lambda g^{(5+p)}_{MN} = 8 \pi  G_{5+p} T_{MN}.
\end{equation}
In these equations we use the following notation for the 5+$p$ coordinates: $X^M\equiv (x^\mu, \theta_i , y)$, where 
$\mu=0,1,2,3$, and $i=1, \dots p$. The four coordinates $x^\mu$ denote to the coordinates that mimic our universe, 
the $\theta_i$'s $\in [0,2\pi]$ denote to the $p$ compact coordinates and the $R_i$'s signal the sizes of the 
corresponding compact dimensions. Finally $y$ denotes the non-compact extra dimension. The superscript 
in the determinant $g^{(5+p)}$ emphasizes the fact that the metric is $(5+p)$D. $G_{5+p}$ is the Newton constant 
in ($5+p$)D and the energy-momentum tensor  
$T_{MN} \equiv \frac{2}{\sqrt{|g^{(5+p)}|}} \frac{\delta S}{\delta g^{MN}}$, corresponds to 
the one produced by the  brane. 

With this setup and appropriate fine-tuning between the brane tension $\sigma$ and the bulk cosmological 
constant $\Lambda$, which are related to $\kappa$ as follows
\begin{equation}\label{tension}
\sigma=\frac{2(3+p)}{8\pi G_{5+p}} \kappa , \quad
\Lambda=-\frac{(3+p)(4+p)}{16\pi G_{5+p}}  \kappa^2 =
-\frac{(4+p) \, \sigma}{4} \kappa,
\end{equation}
there exists a solution to the (5+$p$)D Einstein equations with metric
\begin{equation}\label{mebra}
ds_{5+p}^{\, 2} = e^{-2 \kappa  |y|}
\left[\eta_{\mu\nu}dx^{\mu}dx^{\nu}-\sum_{i=1}^pR^2_id\theta_i^2
\right]-dy^2.
\end{equation}
Here $\eta_{\mu\nu}$ is the 4D Minkowski metric  and without loss of generality it was assumed that the brane is 
at the position $y=0$. At $y=$constant we have 4D flat hypersurfaces extended by $p$ compact extra dimensions.

The interest in these setups comes from its property of 
localizing on the brane: scalar, gauge and gravity fields due to the gravity produced by the brane itself. We 
emphasize that this property is valid whenever there are $p$ extra compact dimensions 
\cite{Dubovsky:2000av,Oda:2000zc}. In the limiting case $p=0$, the model localizes scalar and gravity fields 
but not gauge fields. A short discussion about the consistency of both the KK and the RS compactifications,  as 
well as a discussion of the moduli fixing mechanisms and stability of the setup can be found for instance 
in \cite{Linares:2011sg}. In the literature there are already different analysis of low energy physics effects in these 
setups such as the electric charge conservation \cite{Dubovsky:2000av}, the Casimir effect between two conductor 
hyperplates \cite{Linares:2007yz,Linares:2008am,Frank:2008dt,Linares:2010uy}, the Liennard-Wiechert potentials, 
the Hydrogen Lamb shift \cite{MoralesTecotl:2006eh} and perturbations to the ground state of the Helium atom 
\cite{Garrido:2012sd} among others.

\section{Low Energy Linearized Einstein Equations }\label{GravitationalCase}

In this section we determine the linearized Einstein equations for the perturbations produced by a static source. In 
analogy with the scalar and gauge fields cases discussed in \cite{Linares:2011sg},  we consider a source with the 
topology of a $p$-dimensional torus sitting on the $(3+p)$ brane, which is seen as a punctual mass from the 
perspective of an observer living in the usual 3D low-energy observable part of the brane. In order to solve 
the equations, we follow closely the technique used in \cite{Gregory:2000rh} where authors studied highly 
energetic particles that leave the 4D brane and propagate into the bulk of the 5D RSII model. The main 
difference of the physical situation discussed here respect to the ones previously reported in the literature 
\cite{Gregory:2000rh,Garriga:1999yh,Giddings:2000mu}, is the inclusion of the $p$ extra  compact dimensions.

\subsection{Linearized Einstein equations}

Our starting point are the $(5+p)$D Einstein equations  (\ref{EqsMotion}). Taking the trace of these equations and 
replacing the value of $R$, we obtain the convenient equivalent form 
\begin{equation}\label{lo}
R_{MN}=8\pi G_{5+p}\left(T_{MN}-\frac{1}{3+p}Tg_{MN}\right) + \frac{2}{3+p} \Lambda g_{MN}. 
\end{equation}
In general the linearized Einstein equations that result from considering metric perturbations 
$h_{MN}$ to a known metric solution  $g_{MN}$
\begin{equation}
ds^2=g_{MN}dx^Mdx^N+h_{MN}dx^Mdx^N,
\end{equation}
and energy-momentum tensor perturbations $\delta T_{MN}$, to the equations of motion (\ref{lo}) are given by
\begin{equation}\label{perturbedEq}
\delta R_{MN}=8\pi G_{5+p}
\left[ \delta T_{MN}-\frac{1}{3+p} \left(h_{MN}T + g_{MN}\delta T \right) \right] +  \frac{2}{3+p} \Lambda h_{MN} ,
\end{equation}
where (see for instance \cite{Wald:1984rg})
\begin{equation}\label{VarRicci}
\delta R_{MN}=-\frac{1}{2}\left[\nabla_M \nabla_N \hat{h}+\nabla^A \nabla_A h_{MN}-
\nabla^A \nabla_M h_{NA}-\nabla^A \nabla_N h_{MA}\right],
\end{equation}
and $\hat{h} \equiv g^{MN} h_{MN}$. Following \cite{Gregory:2000rh} we will work in Gaussian Normal (GN) 
coordinates. In such a frame one has
\begin{equation}
h_{yy} = h_{y\bar M}=0,
\end{equation}
where the coordinates  $X^{\bar M}$ label the coordinates of the 4D flat brane and the compact dimensions:
$X^{\bar M} \equiv \{ x^\mu,R_i \theta_i\}$. Accordingly the linearized theory is described by the metric
\begin{equation}
ds^2 = a^2(y)\eta_{\bar M\bar N}dx^{\bar M}dx^{\bar N} + h_{\bar M\bar N}dx^{\bar M}dx^{\bar N} - dy^2 ,
\end{equation}
where $\eta_{\bar M\bar N}=\text{diag}(1,-1,\dots,-1)$ is a $(4+p)$D flat metric. We have also introduced 
the shorthand notation $a(y)\equiv e^{-k|y|}$. It is clear that for the metric of  the RSII$p$ setup
$\hat{h}$ is given simply by $\hat{h}=a^{-2}\eta^{\bar M \bar N}h_{\bar M \bar N} \equiv a^{-2} h$. 

As for the perturbation of the energy-momentum tensor, we shall consider a static source at the position $y_0$ with 
the topology of a $p$D torus, i.e., we consider that the massive object is located a distance $y_0$ away from the 
brane. From these considerations it is clear that during the computation, the perturbed energy-momentum tensor 
resides entirely on the bulk  and is given by
\begin{equation}\label{DefT}
\delta T^{MN}=\frac{m^{(5+p)}}{\sqrt{|g^{(5+p)}|}}\frac{dx^M}{ds}\frac{dx^N}{ds}\delta^3(\vec{x} - 
\vec{x}_0 )\delta(y-y_0),
\end{equation}
where $\frac{dx^M}{ds}= (1,\vec{0})$. A technicality of our calculation is that if $y_0>0$ in (\ref{DefT}), it means we 
are considering an energy-momentum tensor residing to the right of the brane, however the RSII$p$ model owns the 
symmetry $z \rightarrow -z$. Then although we will work entirely only to the right of the brane it should be understood 
that matter is symmetric with respect to the brane and therefore there exists another source located at position 
$-y_0$. The two symmetrical located sources together with the fact that we are considering only symmetric 
perturbations to the metric justify the way in which the computation is done \cite{Gregory:2000rh}. 
Because we are interested in the gravitational potential produced by a source placed on the brane, so after 
computing the solution to the linearized equations we will consider the limit $y_0 \rightarrow 0$,
and the perturbations will appear as generated by a source of mass $M=2m^{(5+p)}$ on the brane. 

It is clear that for an energy-momentum tensor on the bulk, the second term on the right hand of the equation 
(\ref{perturbedEq}) vanishes and the third term becomes: 
$\delta T = a^{-2} \eta^{MN} \delta T_{MN} \equiv a^{-2} \delta T^0_{0}$.
Under these considerations the non vanishing linearized Einstein equations on the bulk are
\begin{eqnarray}
\delta R_{yy}&=&8\pi G_{5+p} \, \frac{1}{3+p} \delta T^0_0 \label{ec.yy},\\
\delta R_{\bar M\bar N}-\frac{2\Lambda}{3+p}h_{\bar M\bar N}&=&
8\pi G_{p+5}\left[\delta T_{\bar M\bar N}-\frac{1}{3+p}\eta_{\bar M\bar N}\delta T^0_0  \right],\label{eqmotion}
\end{eqnarray}
where the variation of the Ricci tensor (\ref{VarRicci}), can be explicitly written as \cite{Langlois:2000ia}
\begin{eqnarray}
\delta R_{yy}&=&-\partial_y\left[\frac{\partial_y h}{2a^2}\right],\label{ec.yy1}\\
\delta R_{\bar M\bar N}&=&\frac{1}{2}\partial_y^2h_{\bar M\bar N}-\frac{p}{2}\kappa\partial_yh_{\bar M\bar N} 
+2\kappa^2h_{\bar M\bar N} - \left(\kappa^2h+\frac{\kappa}{2}\partial_yh\right)\eta_{\bar M\bar N}\nonumber\\
&&+ \frac{1}{2a^2}\left(
\partial^{\bar L}\partial_{\bar M}h_{\bar N\bar L}+
\partial^{\bar L}\partial_{\bar N}h_{\bar M\bar L}-\partial^{\bar L}\partial_{\bar L}h_{\bar M\bar N}-
\partial_{\bar N}\partial_{\bar M}h\right). \label{ec.r4d}
\end{eqnarray}
Notice that the role of the $p$ compact extra dimensions at the level of the variation of the Ricci tensor is given 
by the second term in the right hand side of the equation (\ref{ec.r4d}). In the case $p=0$, we recover the 
expression of the variation of the Ricci tensor for the standard RS model \cite{Gregory:2000rh,Langlois:2000ia}. 

\subsection{The perturbation in modes}

In order to solve the linearized Einstein equations, we start solving equation  (\ref{eqmotion}) by inserting 
(\ref{ec.r4d}) into it 
\begin{eqnarray}
&&\frac{1}{2}h''_{\bar M\bar N}-\frac{p}{2}\kappa h'_{\bar M\bar N}+\frac{1}{2a^2}\left(
\partial^{\bar L}\partial_{\bar M}h_{\bar N\bar L}+\partial^{\bar L}\partial_{\bar N}h_{\bar M\bar L} - 
\partial^{\bar L}\partial_{\bar L}h_{\bar M\bar N}-h_{,\bar M\bar N}\right)+2\kappa^2h_{\bar M\bar N}\nonumber\\
&&-(4+p)\kappa^2h_{\bar M\bar N}=8\pi G_{p+5}\left[\delta T_{\bar M\bar N}-\frac{1}{3+p}
\eta_{\bar M\bar N}\delta T^0_0\right] +\left(\kappa^2h+\frac{\kappa}{2}h'\right)\eta_{\bar M\bar N}, \label{eqmotion2}
\end{eqnarray}
where the prime denotes the derivative with respect to the $y$ coordinate. At his point it is convenient to introduce a 
consideration about the modes spectrum of the metric perturbations into the equation, dictated by the geometry 
of the setup. Formally we write down the metric perturbation in a Fourier series expansion due to the compact 
coordinates 
\begin{equation}
h_{\bar M\bar N}(x,\theta_i,y) = \prod_{k=1}^p \frac{1}{\sqrt{2\pi R_k}} \sum_{\vec{n}} 
(h_{\bar M\bar N}(x,y))_{(\vec{n})}  e^{i \vec{n}\cdot \vec \theta},
\end{equation}
where $\vec{n}$ denotes the collection of $p$ different indexes $\vec{n} = (n_1,n_2,\dots ,n_p)$ taking values in 
$\mathbb{Z}$, $\vec{\theta}$ is a $p$ dimensional vector whose components are the $p$ compact coordinates 
$\theta_k:  \vec{\theta}= (\theta_1, \theta_2, \dots, \theta_p)$ and $\sum_{\vec{n}}$ is the collection of $p$ sums 
$\sum_{\vec{n}}= \sum_{n_1=-\infty}^\infty \cdots  \sum_{n_p=-\infty}^\infty$. The functions
$e^{i \vec{n}\cdot \vec \theta}$ correspond to the basis of the Fourier decomposition along the compact directions. It 
is well known that 
toroidal dimensional reductions a la Kaluza-Klein lead to consistent  lower dimensional theories (see e.g., 
\cite{Appelquist:1987nr} and references therein)
which although do not come with a mechanism to fix the radii of the $T^p$ torus, invoking agreement with 
phenomenology at enough low energies, in particular agreement with the value of the electron charge,  it is 
possible to set a bound to the radius of the order Planck length
\cite{Klein:1926tv}. In the following we shall consider a low energy approximation so that we truncate the massive 
KK modes of the compact dimensions but keeping those that correspond to the noncompact dimension (so far 
encoded in the $y$ dependence of $h_{MN}$), meaning that 
we assume the scale energy of the former is much smaller than that of the latter. Under these considerations 
we are performing the dimensional reduction on the $T^p$ torus or equivalently we are keeping only the zero 
mode of the Fourier expansion, i.e. 
\begin{equation}
h_{\bar M\bar N}(x,\theta_i,y) \approx (h_{\bar M\bar N}(x,y))_{\vec{0}}\, .
\end{equation}
>From here onwards we replace in equations (\ref{ec.yy}) and (\ref{eqmotion}), the whole 
metric perturbation by its zero mode.

Under this consideration the laplacian operator simplifies to:
$\partial^{\bar L}\partial_{\bar L}=\Box+{\partial^{\theta_i}\partial_{\theta_i}} =\Box $, and equation (\ref{eqmotion2})
can be rewritten as 
\begin{eqnarray}
&&\frac{1}{2}h''_{\bar M\bar N}-\frac{p}{2}\kappa h'_{\bar M\bar N}+
\frac{1}{2a^2}\left(\partial^{\bar L}\partial_{\bar M}h_{\bar N\bar L}+\partial^{\bar L}\partial_{\bar N}h_{\bar M\bar L}-
\Box h_{\bar M\bar N}-h_{,\bar M\bar N}\right)+2\kappa^2h_{\bar M\bar N}\nonumber\\
&&-(4+p)\kappa^2h_{\bar M\bar N} = 
8\pi G_{p+5}\left[\delta T_{\bar M\bar N}-\frac{1}{3+p}\eta_{\bar M\bar N}\delta T^0_0 \right]+
\left(\kappa^2h+\frac{\kappa}{2}h'\right)\eta_{\bar M\bar N}.
\label{ec.4d1}
\end{eqnarray}
Introducing the shorthand definition
\begin{equation}\label{eqxi}
\xi_{\bar M}=h^{\bar L}_{\phantom{\bar L}\bar M,\bar L}-\frac{1}{2}h_{,\bar M},
\end{equation}
equation  (\ref{ec.4d1}) takes the form
\begin{eqnarray}
\frac{1}{2}h''_{\bar M\bar N}-\frac{p}{2}\kappa h'_{\bar M\bar N}-\frac{1}{2a^2}\Box h_{\bar M\bar N}
-(2+p)\kappa^2h_{\bar M\bar N}= &&  \nonumber \\
8\pi G_{p+5}\left[\delta T_{\bar M\bar N}-\frac{1}{3+p} \eta_{\bar M\bar N}  T^0_0 \right] 
+\left(\kappa^2h+\frac{\kappa}{2}h'\right) \eta_{\bar M\bar N}&-&\frac{1}{2a^2}
\left(\xi_{\bar M,\bar N}+\xi_{\bar N,\bar M}\right). 
\label{ec.4d2}
\end{eqnarray}
We can consider the following gauge transformation
\begin{equation}
h_{\bar M\bar N}=\bar h_{\bar M\bar N}+u_{\bar M,\bar N}+u_{\bar N,\bar M},
\end{equation}
where $u_{\mu}$ satisfies
\begin{equation}
u_{\bar M}''-p\kappa u_{\bar M}'-2(2+p)\kappa^2u_{\bar M}-\frac{1}{a^2}\Box u_{\bar M}=-\frac{1}{a^2}\xi_{\bar M}.
\end{equation}
It follows then  that  $\bar h_{\bar M\bar N}$ should satisfy
\begin{eqnarray}
\frac{1}{2}\bar h''_{\bar M\bar N}-\frac{p}{2}\kappa \bar h'_{\bar M\bar N}-\frac{1}{2a^2}\Box \bar h_{\bar M\bar N}
-(2+p)\kappa^2\bar h_{\bar M\bar N} &=& \nonumber \\
8\pi G_{p+5}\left[\delta T_{\bar M\bar N}-\frac{1}{3+p} \eta_{\bar M\bar N}\delta T^0_0 \right] &+& 
\left(\kappa^2h+\frac{\kappa}{2}h'\right) \eta_{\bar M\bar N}.
\label{ec.4d3}
\end{eqnarray}
The strategy to solve this equation is the following. We can think the right hand side of the equation (\ref{ec.4d3}) 
as an effective energy-momentum tensor $T_{\bar{M} \bar{N}}^{eff}$, in such a way that 
\begin{equation}\label{DefTeff}
8\pi G_{p+5}\left[\delta T_{\bar M\bar N}-\frac{1}{3+p} \eta_{\bar M\bar N}\delta T^0_0 \right] + 
\left(\kappa^2h+\frac{\kappa}{2}h'\right) \eta_{\bar M\bar N} \equiv  8\pi G_{p+5} T_{\bar{M} \bar{N}}^{eff}.
\end{equation}
Therefore the equation (\ref{ec.4d3}) takes the form
\begin{equation}\label{InhomEq}
\frac{1}{2}\bar h''_{\bar M\bar N}-\frac{p}{2}\kappa \bar h'_{\bar M\bar N}-\frac{1}{2a^2}\Box \bar h_{\bar M\bar N}
-(2+p)\kappa^2\bar h_{\bar M\bar N} = 8\pi G_{p+5} T_{\bar{M} \bar{N}}^{eff}.
\end{equation}
Solving this equation requires to know the solutions to the homogeneous equations, once we have these solutions 
we can compute the Green function and with it solving the inhomogeneous equation (\ref{InhomEq}).  It is 
also convenient a this point to expand $\bar{h}_{MN}(x,y)$ in terms of the functions $\psi_m(y)$, which correspond
to the modes structure of the metric perturbations due to the 
non-compact dimension $y$
\begin{equation}
(h_{\bar M\bar N}(x,y))_{\vec{0}} = \left( \int  (h_{\bar M\bar N}(x))_m \psi_m(y) \, dm \right)_{(\vec{0})}.
\end{equation}
Plugging this ansatze in the left hand side of equation (\ref{InhomEq}), allows us to perform a separation of 
variables in the differential operator. Introducing the separation constant $m$ lead us to have an 
equation for $\psi_m(y)$ of the following form
\begin{equation}
\left(\partial_y^2-p\kappa \partial_y-2(2+p)\kappa^2+\frac{m^2}{a^2}\right)\psi_m(y)=0.
\end{equation}
This equation can be rewritten as a Bessel equation. In order to do that we perform the variable change
$\xi(y)= \epsilon a^{-1}(y)$, and we introduce the rescaled function $\psi(\xi)=\xi^{p/2}\tilde\psi(\xi)$, obtaining
\begin{equation}
\left[ \partial_{\xi}^2+\frac{1}{\xi}\partial_{\xi}+m^2-\frac{\alpha^2}{\xi^2} \right]\tilde\psi=0,
\end{equation}
where the constant $\alpha \equiv 2+\frac{p}{2}$, contains the information about the number 
of extra compact dimensions. 

For the massless mode ($m=0$) the solution is
\begin{equation}\label{ec.mocerog}
\tilde\psi_0(\xi)=A_1\xi^{\alpha}+A_2\xi^{-\alpha} \hspace{0.5cm} \Rightarrow \hspace{0.5cm} 
\psi_0(\xi)=  a_1\xi^{p+2}+a_2\xi^{-2},
\end{equation}
where $a_i $ are integration constants. We take $a_1=0$ in order to have a normalizable solution, which 
is explicitly given by
\begin{equation}
\psi_0(y)=\sqrt{\left(1+\frac{p}{2}\right)\kappa}\,e^{-2\kappa y}.
\end{equation}
For the massive modes ($m>0$) we obtain
\begin{equation}\label{MassiveMode}
\psi_m(y)=e^{\frac{p\kappa y}{2}}\sqrt{\frac{m}{2 \kappa}}\left[a_mJ_{\alpha} \left(\frac{m}{\kappa} e^{\kappa y}\right)
+b_mN_{\alpha} \left(\frac{m}{\kappa} e^{\kappa y}\right) \right],
\end{equation}
where the constants  $a_m$ and $b_m$  are given by
\begin{equation}
a_m=-\frac{A_m}{\sqrt{1+A_m^2}},\quad b_m=\frac{1}{\sqrt{1+A_m^2}} ,
\end{equation}
with
\begin{equation}
A_m=\frac{N_{\alpha-1}\left( \frac{m}{\kappa} \right)-\frac{2\kappa}{m}N_{\alpha}\left(  \frac{m}{\kappa}  \right)}
{J_{\alpha-1}\left(  \frac{m}{\kappa}  \right)-\frac{2\kappa}{m}J_{\alpha}\left(  \frac{m}{\kappa}  \right)}. 
\end{equation}
In order to simplify further this expression it is convenient to take the approximation of light modes $m\ll \kappa^{-1}$, 
this is plausible because these are the modes contributing the most to the potential. In this approximation  
\begin{equation}
A_m=\frac{\Gamma(\alpha-1)\Gamma(\alpha)}{\pi}\left(\frac{m}{2 \kappa}\right)^{2-2\alpha} ,
\end{equation}
and therefore the coefficients $a_m$ and $b_m$ are given by
\begin{equation}
a_m=-1, \hspace{0.5cm}
b_m=\frac{\pi}{\Gamma(\alpha-1)\Gamma(\alpha)}\left(\frac{m}{2\kappa}\right)^{2\alpha-2}.
\end{equation}
Plugging these coefficients into equation (\ref{MassiveMode}) and considering the same light modes approximation 
in the Bessel and Neumman functions we get
\begin{eqnarray}
\psi_m(0)&=&\sqrt{\frac{m}{2 \kappa}}\frac{1}{\Gamma(\alpha-1)}\left(\frac{m}{2 \kappa}\right)^{\alpha-2},\\
\psi_m(y')&=&
-e^{\frac{p}{2} \kappa y'} \sqrt{\frac{m}{2\kappa}}J_{\alpha}\left(\frac{m}{\kappa} e^{\kappa y' } \right).
\end{eqnarray}
Notice we are computing the massive modes in two different points of the $y$ coordinate because with these 
functions we are constructing the two points Green function.

\subsection{The Green function}\label{SecGreen}

With the eigenfunctions $\psi_m(y)$ it is straightforward to construct the Green function
\begin{eqnarray}
G_R(x,x',y=0,y')&=&-\frac{\psi_0(0)\psi_0(y')}{4\pi r}-\int_0^{\infty} dm \, \psi_m(0)\psi_m(y') \frac{e^{-mr}}{4\pi r} \\
&=&-\frac{1}{4\pi r}\left(1+\frac{p}{2}\right)\frac{1}{\kappa \, \xi^2}
+\frac{\xi^{\frac{p}{2}}}{\Gamma(\alpha-1)2^{\alpha-1} \kappa^{1-\alpha + \frac{p}{2}}}
\int_0^{\infty}dm\,
m^{\alpha}J_{\alpha}\left(m\xi\right)\frac{e^{-mr}}{m}. \nonumber
\end{eqnarray}
Explicit evaluation of this function depends on the parity of the number $p$ of compact dimensions. 

\subsubsection{$p$ odd:}

For this case we have that $\alpha$ takes semi-integer values and the Green function is

\begin{eqnarray}
G_R(x,x',y=0,y')=&&\frac{1}{4\pi r}\sqrt{\frac{2}{\pi}}
\frac{(-1)^{\alpha-\frac{1}{2}}  \xi^{\alpha+\frac{p}{2}}}{\Gamma(\alpha-1)2^{\alpha-1} \kappa^{1-\alpha + \frac{p}{2}} }
\left(\frac{d}{\xi d\xi}\right)^{\alpha-\frac{1}{2}}
\left[\frac{\pi}{2\xi}-\frac{\arctan\left(\frac{r}{\xi}\right)}{\xi}
\right]\nonumber\\
&&-\frac{1}{4\pi r}\left(1+\frac{p}{2}\right)\frac{1}{\kappa \, \xi^2}.
\end{eqnarray}
Using the relation
\begin{equation}
\left(\frac{d}{\xi d\xi}\right)^{\alpha-\frac{1}{2}}
\left[\frac{1}{\xi}\right]=\frac{(-1)^{\alpha-\frac{1}{2}}\left[2\left(\alpha-1\right)\right]! 
(\alpha-1)!}{2^{\alpha-\frac{3}{2}}2^{-2\alpha+2}\sqrt{\pi}\xi^{2\alpha}\left[2\left(\alpha-1\right)\right]!}=
\frac{(-1)^{\alpha-\frac{1}{2}}(\alpha-1)!}{2^{-\alpha+\frac{1}{2}}\sqrt{\pi}\xi^{2\alpha}},
\end{equation}
the Green function can be written as
\begin{equation}
G_R(x,x',y=0,y')=-
\frac{1}{4\pi r}\sqrt{\frac{2}{\pi}}
\frac{(-1)^{\alpha-\frac{1}{2}}\xi^{\alpha+\frac{p}{2}}}{\Gamma(\alpha-1)2^{\alpha-1} \kappa}
\left(\frac{d}{\xi d\xi}\right)^{\alpha-\frac{1}{2}}
\left[\frac{\arctan\left(\frac{r}{\xi}\right)}{\xi} \right].
\end{equation}
The derivative can be evaluated, recalling the relation
\begin{equation}
\frac{d}{\xi d\xi}f\left(\xi\right)=2\left.\frac{d}{d\beta}f\left(\sqrt{\xi^2+\beta}\right)\right|_{\beta=0} ,
\end{equation}
which leads to the final form of the Green function
\begin{eqnarray}
G_R(x,x',y=0,y')&=&-
\frac{1}{4\pi r}\sqrt{\frac{2}{\pi}}
\frac{(-1)^{\alpha-\frac{1}{2}} \xi^{\alpha+\frac{p}{2}}}{\Gamma(\alpha-1)2^{\alpha-1} \kappa }
2^{\alpha-\frac{1}{2}}\nonumber\\
&&\left[
-\Gamma\left(\alpha+\frac{1}{2}\right) 
\frac{r(-1)^{\alpha-\frac{1}{2}}}{2\alpha\left(r^2+\xi^2\right)^{\alpha+\frac{1}{2}}}
F\left(1,\alpha+\frac{1}{2};\alpha+1;{\frac {\xi^2}{r^2+\xi^2}} \right)\right. \label{GreenOdd}\\
&+& \left. \frac{(-1)^{\alpha-\frac{1}{2}}\Gamma(\alpha)}{\sqrt{\pi}}\frac{1}{\xi^{2\alpha}} 
\arcsin\left(\frac{\xi}{\sqrt{r^2+\xi^2}}\right) +\frac{(-1)^{\alpha-\frac{1}{2}}}{\sqrt{\pi}\xi^{2\alpha}}
\Gamma(\alpha)\arctan\left(\frac{r}{\xi}\right)
\right].\nonumber
\end{eqnarray}

\subsubsection{$p$ even:}

For even $p$, $\alpha$ takes integer values and the Green function is
\begin{eqnarray}
G_R(x,x',y=0,y') &=& -\frac{1}{4\pi r}\left(1+\frac{p}{2}\right)\frac{1}{\kappa \, \xi^2} +  \\
& & -\frac{1}{4\pi r}
\frac{(-1)^{\alpha} \xi^{\alpha+\frac{p}{2}}}{\Gamma(\alpha-1)2^{\alpha-1} \kappa^{1-\alpha + \frac{p}{2}}}
\left(\frac{d}{\xi d\xi}\right)^{\alpha-1} \left[\frac{1}{\xi^2}-\frac{r}{\xi^2\sqrt{r^2+\xi^2}} \right]. \nonumber
\end{eqnarray}
In a similar way as the former case, we obtain finally the Green function  for even compact dimensions
\begin{equation}
G_R(x,x',y=0,y')= \frac{1}{4\pi r}
\frac{(-1)^{\alpha}\xi^{\alpha+\frac{p}{2}}}{\Gamma(\alpha-1)2^{\alpha-1} \kappa}
\left(\frac{d}{\xi d\xi}\right)^{\alpha-1}
\left[\frac{r}{\xi^2\sqrt{r^2+\xi^2}}
\right].
\end{equation}  

\section{Solutions}\label{solutions}

We are now in position to compute the solutions to the linearized Einstein equations in the  low energy  regime. 
The order in which the solutions are obtained is the following.  We start solving 
the equation  (\ref{ec.yy}) where the Ricci tensor is given by equation (\ref{ec.yy1}). This happen  
because we have to know the expression for the combination: $\kappa^2h(x',y')+\frac{\kappa}{2}\partial_yh(x',y')$, 
in order to solve for the perturbations $\bar{h}_{MN}$ of the equations (\ref{eqmotion}).

\subsection{Solution of the $yy$ equation}

We start integrating twice equation (\ref{ec.yy}) 
\begin{equation}
-\partial_y\left[\frac{\partial_y h}{2a^2}\right] =8\pi G_{5+p} \, \frac{1}{3+p} \delta T^0_0.
\end{equation}
After the first integral we directly get
\begin{equation}
h'=- 2a^2  \, \frac{8\pi G_{5+p}}{3+p}  \int_y^{\infty} dy \,  \delta T^0_0 + 2a^2C(x),
\end{equation}
and after the second integral we obtain
\begin{equation}
h=-\int_y^{\infty} dy \left[ 2a^2  \, \frac{8\pi G_{5+p}}{3+p}  \int_y^{\infty} dz \,  \delta T^0_0(z) - 2a^2C(x) \right] +D(x),
\end{equation}
here $C(x)$ y $D(x)$ are functions to be determined. From the explicit form of $a(y)$ we can evaluate in a 
straightforward way, the second term of the integral in the equation above
\begin{equation}
\int_y^{\infty} dy \, 2a^2(y)= \frac{a^2(y)}{\kappa},
\end{equation}
whereas for the first term we use an integration by parts
\begin{equation}
\int_y^{\infty} dy \, 2a^2\int_y^{\infty} dz \, \delta T^0_0(z) =
\frac{a^2}{\kappa}\int_y^{\infty} dy \, \delta T^0_0 -\int_y^{\infty} dy \frac{a^2}{\kappa} \delta T^0_0,
\end{equation}
obtaining that $h(y)$ is of the form
\begin{equation}
h=-\frac{8\pi G_{5+p}}{(3+p)\kappa}\left[a^2\int_y^{\infty} dy \, \delta T^0_0 -\int_y^{\infty} dy \, a^2 \delta T^0_0 \right] +
\frac{a^2}{\kappa} C(x) +D(x).  \label{ec.traza}
\end{equation}
So far we have only considered the perturbation in the bulk. The role played by the brane in the solution 
appears through the junction conditions
\begin{equation}\label{junction}
K_{\bar M\bar N}=-\frac{8\pi G_{5+p}}{2}\left(S_{\bar M\bar N}-\frac{1}{3+p}\eta_{\bar M\bar N}\, a^2 S\right),
\end{equation}
which constitute a connection between the metric perturbations living in the bulk and the matter perturbations 
confined to the brane ($S_{\bar{M} \bar{N}}$) \cite{Israel:1966rt}. In a GN coordinate system, the extrinsic curvature 
is given by the simple expression
\begin{equation}\label{DefK}
\quad K_{\bar M\bar N}=\frac{1}{2}\partial_y\left(a^2\eta_{\bar M\bar N}+h_{\bar M\bar N}\right),
\end{equation}
whereas the energy-momentum tensor on the brane is given by
\begin{equation}\label{DefS}
 S_{\bar M\bar N} = -\sigma\left(a^2\eta_{\bar M\bar N}+h_{\bar M\bar N}\right)+\delta T_{\bar M\bar N}. 
\end{equation}
In equation (\ref{junction}) we are using the definition $S \equiv a^{-2} \eta^{\bar{M} \bar{N}} S_{\bar M\bar N}$. 
Plugging in the expressions (\ref{DefK}) and (\ref{DefS}) in the equation  (\ref{junction}) and considering the energy 
momentum tensor (\ref{DefT}) and the relation between the brane tension and the adS radius (\ref{tension}), 
we obtain after taking the trace of the junction condition that
\begin{equation}
\left.\partial_yh+2\kappa h=\frac{8\pi G_{5+p}}{3+p}\delta T\right|_{y=0}=
\left.\frac{4\pi G_{5+p}\kappa}{3+p}\frac{m^{(5+p)}}{a^{2+p}(y')}\delta(y-y_0)\delta^3(\vec{x}-\vec{x}_0)\right|_{y=0}.
 \label{eccb}
\end{equation}
This means that the points  $y=0$ and  $y_0$  never coincide and therefore $\delta(y'-y_0)$ is null. Substitution of 
expression  (\ref{ec.traza}) into Eq. (\ref{eccb}), allows us to find the expression for the function $D(x)$ which is
given by the equation
\begin{equation}
2\kappa D(x)+16\pi G_{5+p}\int_0^{\infty}a^2(y')\delta T(y')dy'=0.
\end{equation}
Once we know the value of $D(x)$, we can evaluate the combination of $h$ and $h'$ that appears in the definition 
(\ref{DefTeff}) of $T^{eff}_{\bar{M} \bar{N}}$
\begin{eqnarray}
\kappa^2h(x',y')+\frac{\kappa}{2}\partial_{y'}h(x',y')&=&
-8\pi G_{5+p}\kappa\int_0^{y'}a^2(z)\delta T(z)dz\nonumber\\
&=&-\frac{8\pi G_{5+p}\kappa m^{(5+p)}}{a^{2+p}(y_0)}\theta(y'-y_0)\delta^3(\vec{x}'-\vec{x}_0). 
\label{combination}
\end{eqnarray}

\subsection{The $\bar{h}_{00}$ component}

Once we have  the solution of the equation (\ref{ec.yy}) and as a consequence the expression for the combination 
(\ref{combination}),  we can compute the expressions for the metric perturbations. Using the Green function of the 
subsection (\ref{SecGreen}) we have that
\begin{eqnarray}
\bar h_{00}(r,y=0)&=&8\pi G_{5+p}\int d^3x'\int dy'G(\vec{x},y=0;\vec{x'},y')
\left[\left(\delta T_{00}(x',y')-\frac{1}{3+p}\eta_{00}\delta T^0_0(x',y')\right)\right.\nonumber\\
&&\left.+
\frac{1}{8\pi G_{5+p}}\left(\kappa^2h(x',y')+\frac{\kappa}{2}\partial_yh(x',y')\right)\eta_{00}\right],
\label{ec.metricabrana00}
\end{eqnarray}
where according to the energy-momentum tensor (\ref{DefT})
\begin{equation}\label{source}
\delta T_{00}(x',y')-\frac{1}{3+p}\eta_{00}\delta T^0_0(x',y') = 
\frac{2+p}{3+p}\frac{m^{(5+p)}}{a^{2+p}(y')}\delta(y'-y_0)\delta^3(\vec{x}'-\vec{x}_0).
\end{equation}  
Plugging in expressions (\ref{combination}) and (\ref{source}) in (\ref{ec.metricabrana00}) we obtain
\begin{eqnarray}
\bar h_{00}(r,y=0)&=& 
8\pi G_{5+p}\frac{2+p}{3+p}\frac{m^{(5+p)}}{a^{2+p}(y_0)}G_R(x,x'=x_0,y=0,y'=y_0)
\nonumber\\
&& - \frac{8\pi G_{5+p}\kappa m^{(5+p)}}{a^{2+p}(y_0)}\int dy'G_R(x,x'=x_0,y=0,y')\theta(y'-y_0).
\label{ec.metricabrana}
\end{eqnarray}
As we have discussed, the explicit form of the Green function depends of the parity of the number of compact 
extra dimensions $p$, and therefore this also happen for the metric component $\bar{h}_{00}$

\subsubsection{$p$ odd}

In the case in which $p$ is odd, we use the Green function (\ref{GreenOdd}) obtaining
\begin{eqnarray}
\bar h_{00}(r,y=0)
&=&-8\pi G_{5+p}\frac{2+p}{3+p}\frac{m^{(5+p)}}{a^{2+p}(y_0)}\frac{1}{4\pi r}\sqrt{\frac{2}{\pi}}
\frac{(-1)^{\alpha-\frac{1}{2}}\epsilon\xi^{\alpha+\frac{p}{2}}}{\Gamma(\alpha-1)2^{\alpha-1}}
\left(\frac{d}{\xi d\xi}\right)^{\alpha-\frac{1}{2}}
\left.\left[\frac{\arctan\left(\frac{r}{\xi}\right)}{\xi}
\right]\right|_{\xi=\xi_0}
\nonumber\\
&&+
\frac{8\pi G_{5+p}\kappa m^{(5+p)}}{a^{2+p}(y_0)}\int \frac{d\xi}{k\xi}\frac{1}{4\pi r}\sqrt{\frac{2}{\pi}}
\frac{(-1)^{\alpha-\frac{1}{2}}\epsilon\xi^{\alpha+\frac{p}{2}}}{\Gamma(\alpha-1)2^{\alpha-1}}
\left(\frac{d}{\xi d\xi}\right)^{\alpha-\frac{1}{2}}
\left[\frac{\arctan\left(\frac{r}{\xi}\right)}{\xi}
\right]\theta(\xi-\xi_0).\nonumber\\
\end{eqnarray}

$\bullet$ Example: $p$=1

In particular, if we take the  value $p=1$, we have
\begin{eqnarray}
G_R(x,x',y=0,y')^{(1)}
&=&-
\frac{1}{4\pi r}
\frac{\epsilon\xi^3}{\pi}
\left(\frac{d}{\xi d\xi}\right)^{2}
\left[\frac{\arctan\left(\frac{r}{\xi}\right)}{\xi}
\right]\nonumber\\
&=&-\frac{1}{4\pi r}\frac{\epsilon}{\pi}\left[\frac{5r}{\xi^3\left(1+\frac{r^2}{\xi^2}\right)}
-\frac{2r^3}{\xi^5\left(1+\frac{r^2}{\xi^2}\right)^2}
+\frac{3}{\xi^2}\arctan\left(\frac{r}{\xi}\right)\right],
\end{eqnarray}
and  $\bar h_{00}$  is given by
\begin{eqnarray}
\bar h_{00}&=&\frac{6\pi G_{6}m^{(6)}}{a^{2}(y_0)}\frac{\epsilon}{4\pi^2 r}
\left[\frac{5r}{\xi_0^3\left(1+\frac{r^2}{\xi_0^2}\right)}
-\frac{2r^3}{\xi_0^5\left(1+\frac{r^2}{\xi_0^2}\right)^2}
+\frac{3}{\xi_0^2}\arctan\left(\frac{r}{\xi_0}\right)\right]
\nonumber\\
&&-\frac{8\pi G_{6}m^{(6)}}{ a^{2}(y_0)}\frac{\epsilon}{4\pi^2 r}
\left.\left[
-\frac{3}{2}\frac{\arctan\left(\frac{r}{\xi}\right)}{\xi^2}-\frac{3}{2}\frac{1}{r\xi}-
\frac{1}{2}\frac{\arctan\left(\frac{\xi}{r}\right)}{r^2}+\frac{1}{r\xi\left(1+\frac{r^2}{\xi^2}\right)}
\right]\right|_{\xi=\xi_0}^{\infty}.
\end{eqnarray}
Taking the limit when  $y_0\to0$, $\xi_0=\epsilon$, we obtain for this component
\begin{eqnarray}
\bar h_{00}&=&-\frac{3 G_{6}m^{(6)}}{2\pi^2}
\left[\frac{5}{\epsilon^2\left(1+\frac{r^2}{\epsilon^2}\right)}
-\frac{2r^2}{\epsilon^4\left(1+\frac{r^2}{\epsilon^2}\right)^2}
+\frac{3}{r\epsilon}\arctan\left(\frac{r}{\epsilon}\right)\right]
\nonumber\\
&&
+\frac{2G_{6}m^{(6)}}{\pi^2}\left[
\frac{3}{2}\frac{\arctan\left(\frac{r}{\epsilon}\right)}{r\epsilon}+\frac{3}{2}\frac{1}{r^2}+
\frac{\epsilon}{2}\frac{\arctan\left(\frac{\epsilon}{r}\right)}{r^3}-\frac{1}{r^2\left(1+\frac{r^2}{\epsilon^2}\right)}-
\frac{1}{4}\frac{\pi\epsilon}{r^3} \right].
\end{eqnarray}
It is illustrative to calculate the short and the long distance limits
\begin{eqnarray}
\bar h_{00}
&=&-\frac{G_{6}m^{(6)}}{\pi^2}\left[\frac{20}{3\epsilon^2}-\frac{44}{5\epsilon^4}r^2+\dots\right],\quad r\to 0,\\
\bar h_{00}
&=&-\frac{G_{6}m^{(6)}}{\pi^2}\left[
\frac{3\pi}{4\epsilon}\frac{1}{r}+\frac{\epsilon\pi}{2r^3}+\dots
\right]\sim- \frac{2G_Nm}{r}\left(1+\frac{2\epsilon^2}{3}\frac{1}{r^2}\right), \quad r\to \infty,
\label{potential1}
\end{eqnarray}
where we have defined the effective 4D Newton constant in terms of the 6D one as
\begin{equation}\label{G4inG6}
\quad G_N=\frac{3G_{(6)}}{8\pi \epsilon}.
\end{equation}

\subsubsection{Example: $p$=2}

For the even case we give as an example the value  $p =2$. In this case the Green function is
\begin{eqnarray}
G_R(x,x',y=0,y')^{(2)}&=&-
\frac{1}{4\pi r}
\frac{\epsilon\xi^4}{4}
\left(\frac{d}{\xi d\xi}\right)^2
\left[\frac{r}{\xi^2\sqrt{r^2+\xi^2}}
\right]\nonumber\\
&=&-\frac{1}{4\pi }
\frac{\epsilon}{4}
\left[
\frac{8}{\xi^2\sqrt{r^2+\xi^2}}+\frac{4}{\left(r^2+\xi^2\right)^{\frac{3}{2}}}
+\frac{3\xi^2}{\left(r^2+\xi^2\right)^{\frac{5}{2}}}
\right],
\end{eqnarray}
and the potential is given by
\begin{eqnarray}
\bar h_{00}^{(2)}
&=&-\left.\frac{4}{5}\frac{8\pi G_{7}m^{(7)}}{a^{3}(y_0)}
\frac{1}{4\pi } \frac{\epsilon}{4}
\left[ \frac{8}{\xi^2\sqrt{r^2+\xi^2}}+\frac{4}{\left(r^2+\xi^2\right)^{\frac{3}{2}}}
+\frac{3\xi^2}{\left(r^2+\xi^2\right)^{\frac{5}{2}}} \right]
\right|_{\xi=\xi_0}\nonumber\\
&&+\frac{8\pi G_{7}m^{(7)}}{\epsilon a^{3}(y_0)}\int_{\xi_0}^{\infty}\frac{\epsilon d\xi}{\xi}
\frac{1}{4\pi }
\frac{\epsilon}{4}
\left[
\frac{8}{\xi^2\sqrt{r^2+\xi^2}}+\frac{4}{\left(r^2+\xi^2\right)^{\frac{3}{2}}}
+\frac{3\xi^2}{\left(r^2+\xi^2\right)^{\frac{5}{2}}}
\right]\theta(\xi-\xi_0).
\end{eqnarray}
Evaluating the integral we finally have
\begin{equation}
\bar h_{00}^{(2)}=-\frac{2G_{7}m^{(7)}}{5}
\left[ \frac{8}{\epsilon\sqrt{r^2+\epsilon^2}}+\frac{4\epsilon}{\left(r^2+\epsilon^2\right)^{\frac{3}{2}}}
+\frac{3\epsilon^3}{\left(r^2+\epsilon^2\right)^{\frac{5}{2}}} \right] +\frac{G_{7}m^{(7)}}{2 }
\frac{4r^2+5\epsilon^2}{\epsilon(\epsilon^2+r^2)^{3/2}}.
\end{equation}
The short and long distance limits for this case are
\begin{eqnarray}
\bar h_{00}^{(2)}&=&- G_{7}m^{(7)}\left[ \frac{7}{2\epsilon^2}-\frac{21}{4\epsilon^4}r^2+\dots \right],\quad r\to 0,\\
\bar h_{00}^{(2)}&=& -G_{7}m^{(7)}\left[ \frac{6}{5\epsilon}\frac{1}{r} +\frac{\epsilon}{2r^3}+\dots
\right]\sim- \frac{2G_Nm}{r}\left(1+\frac{5\epsilon^2}{12}\frac{1}{r^2}\right), \quad  r\to \infty, \label{potential2}
\end{eqnarray}
where the 4D Newton constant is
\begin{equation}\label{G4inG7}
G_N=\frac{3G_{(7)}}{5 \epsilon}.
\end{equation}

\subsection{The $\bar{h}_{ij}$ components}

For the spatial components of the induced metric on the brane we proceed as before. In this case
the Green function of the subsection (\ref{SecGreen}) reads
\begin{eqnarray}
\bar h_{ij}(r,y=0)&=&8\pi G_{5+p}\int d^3x'\int dy'G(\vec{x},y=0;\vec{x'},y')
\left[\left(\delta T_{ij}(x',y')-\frac{1}{3+p}\eta_{ij}\delta T^0_0(x',y')\right)\right.\nonumber\\
&&\left.+ \frac{1}{8\pi G_{5+p}}\left(\kappa^2h(x',y')+\frac{\kappa}{2}\partial_yh(x',y')\right)\eta_{ij}\right],
\label{ec.metricabrana}
\end{eqnarray}
where this time, according to (\ref{DefT})
\begin{equation}
{\delta T_{ij}(x',y')}-\frac{1}{3+p}\eta_{ij}\delta T(x',y')=
-\frac{\eta_{ij}}{3+p}\frac{m^{(5+p)}}{a^{1+p}(y')}\delta(y'-y_0)\delta^3(\vec{x}'-\vec{x}_0).
\end{equation}
Thus in this case we have in general that
\begin{eqnarray}
\bar h_{ij}(r,y=0)&=&-8\pi G_{5+p}\frac{\eta_{ij}}{3+p}\frac{m^{(5+p)}}{a^{2+p}(y_0)}G_R(x,x'=x_0,y=0,y'=y_0)
\nonumber\\
&&-\frac{8\pi G_{5+p}\kappa m^{(5+p)}}{a^{2+p}(y_0)} \eta_{ij} \int dy'G_R(x,x'=x_0,y=0,y')\theta(y'-y_0).
\label{ec.metricabrana}
\end{eqnarray}
Again the computations have to be worked out in two separate cases depending on the parity of the number 
of extra compact dimensions

\subsubsection{$p$ odd}

This case correspond to have integer values of the parameter $\alpha$, so the expression of the components 
$\bar{h}_{ij}$ is given by 
\begin{eqnarray}
\bar h_{ij}(r,y=0)
&=&8\pi G_{5+p}\frac{\eta_{ij}}{3+p}\frac{m^{(5+p)}}{a^{2+p}(y_0)}\frac{1}{4\pi r}\sqrt{\frac{2}{\pi}}
\frac{(-1)^{\alpha-\frac{1}{2}}\epsilon\xi^{\alpha+\frac{p}{2}}}{\Gamma(\alpha-1)2^{\alpha-1}}
\left(\frac{d}{\xi d\xi}\right)^{\alpha-\frac{1}{2}}
\left.\left[\frac{\arctan\left(\frac{r}{\xi}\right)}{\xi}
\right]\right|_{\xi=\xi_0}
\nonumber\\
&&+
\frac{8\pi G_{5+p}\kappa m^{(5+p)}\eta_{ij}}{a^{2+p}(y_0)}\int dy'\frac{1}{4\pi r}\sqrt{\frac{2}{\pi}}
\frac{(-1)^{\alpha-\frac{1}{2}}\epsilon\xi^{\alpha+\frac{p}{2}}}{\Gamma(\alpha-1)2^{\alpha-1}}
\left(\frac{d}{\xi d\xi}\right)^{\alpha-\frac{1}{2}}
\left[\frac{\arctan\left(\frac{r}{\xi}\right)}{\xi}
\right]\theta(y'-y_0).\nonumber
\end{eqnarray}

$\bullet$ Example $p=1$

Evaluating the Green function for this case lead us to the expression 
\begin{eqnarray}
G_R(x,x',y=0,y')^{(1)} &=&- \frac{1}{4\pi r} \frac{\epsilon\xi^3}{\pi} \left(\frac{d}{\xi d\xi}\right)^{2}
\left[\frac{\arctan\left(\frac{r}{\xi}\right)}{\xi} \right]\nonumber\\
&=&-\frac{1}{4\pi r}\frac{\epsilon}{\pi}\left[\frac{5r}{\xi^3\left(1+\frac{r^2}{\xi^2}\right)}
-\frac{2r^3}{\xi^5\left(1+\frac{r^2}{\xi^2}\right)^2}
+\frac{3}{\xi^2}\arctan\left(\frac{r}{\xi}\right)\right],
\end{eqnarray}
hence  $\bar h_{ij}$ is after taking the limit $y_0\to0$ 
\begin{eqnarray}
\bar h_{ij} &=&\frac{G_{6}m^{(6)}\eta_{ij}}{2\pi^2} \left[\frac{5}{\epsilon^2\left(1+\frac{r^2}{\epsilon^2}\right)}
-\frac{2r^2}{\epsilon^4\left(1+\frac{r^2}{\epsilon^2}\right)^2}
+\frac{3}{r\epsilon}\arctan\left(\frac{r}{\epsilon}\right)\right] \nonumber\\
&& +\frac{2G_{6}m^{(6)}\eta_{ij}}{\pi^2}\left[
\frac{3}{2}\frac{\arctan\left(\frac{r}{\epsilon}\right)}{r\epsilon}+\frac{3}{2}\frac{1}{r^2}+
\frac{\epsilon}{2}\frac{\arctan\left(\frac{\epsilon}{r}\right)}{r^3}-
\frac{1}{r^2\left(1+\frac{r^2}{\epsilon^2}\right)}-\frac{1}{4}\frac{\pi\epsilon}{r^3} \right].
\end{eqnarray}
For astrophysical applications is convenient to calculate the long distantces limit
\begin{eqnarray}
\bar h_{ij}
&=&-\frac{G_{6}m^{(6)}\eta_{ij}}{\pi^2}\left[
-\frac{9\pi}{4\epsilon}\frac{1}{r}+\frac{\epsilon\pi}{2r^3}+\dots
\right]\sim- \frac{2G_Nm}{r}\left(-3+\frac{2\epsilon^2}{3}\frac{1}{r^2}\right)\eta_{ij},\quad  r\to \infty,
\label{potential3}
\end{eqnarray}
where the Newton constant is the same as in equation (\ref{G4inG6}).

\subsubsection{$p=2$}

In this case $\alpha$ is a semi-integer number and the Green function is given by
\begin{eqnarray}
G_R(x,x',y=0,y')^{(2)}&=&-
\frac{1}{4\pi r} \frac{\epsilon\xi^4}{4} \left(\frac{d}{\xi d\xi}\right)^2 \left[\frac{r}{\xi^2\sqrt{r^2+\xi^2}} \right]\nonumber\\
&=&-\frac{1}{4\pi } \frac{\epsilon}{4}
\left[\frac{8}{\xi^2\sqrt{r^2+\xi^2}}+\frac{4}{\left(r^2+\xi^2\right)^{\frac{3}{2}}}
+\frac{3\xi^2}{\left(r^2+\xi^2\right)^{\frac{5}{2}}}\right],
\end{eqnarray}
thus the potential is written as
\begin{eqnarray}
\bar h_{ij}^{(2)}&=&\left.\frac{G_{7}m^{(7)}\epsilon\eta_{ij}}{10a^{3}(y_0)}
\left[
\frac{8}{\xi^2\sqrt{r^2+\xi^2}}+\frac{4}{\left(r^2+\xi^2\right)^{\frac{3}{2}}}
+\frac{3\xi^2}{\left(r^2+\xi^2\right)^{\frac{5}{2}}}
\right]
\right|_{\xi=\xi_0}\nonumber\\
&&+\frac{G_{7}m^{(7)}\epsilon \eta_{ij}}{2 a^{3}(y_0)}
\left.\left[
-\frac{4r^2+5\xi^2}{\xi^2(\xi^2+r^2)^{3/2}}
\right]\right|_{\xi=\xi_0}^{\infty}.
\end{eqnarray}
Evaluating the limits explicitly  we have
\begin{eqnarray}
\bar h_{ij}^{(2)}&=&\frac{G_{7}m^{(7)}\eta_{ij}}{10}
\left[
\frac{8}{\epsilon\sqrt{r^2+\epsilon^2}}+\frac{4\epsilon}{\left(r^2+\epsilon^2\right)^{\frac{3}{2}}}
+\frac{3\epsilon^3}{\left(r^2+\epsilon^2\right)^{\frac{5}{2}}}
\right]
\nonumber\\
&&+\frac{G_{7}m^{(7)}\eta_{ij}}{2 }
\frac{4r^2+5\epsilon^2}{\epsilon(\epsilon^2+r^2)^{3/2}}.
\end{eqnarray}
Taking the large distances limit ($r\to \infty$) we obtain
\begin{eqnarray}
\bar h_{ij}^{(2)}&=&-
G_{7}m^{(7)}\eta_{ij}\left[
-\frac{14}{5\epsilon}\frac{1}{r}
+\frac{\epsilon}{2r^3}+\dots
\right]\sim- \frac{2G_Nm}{r} \left(-\frac{7}{3}+\frac{5\epsilon^2}{12}\frac{1}{r^2}\right)\eta_{ij},
 \label{potential4}
\end{eqnarray}
where $G_N$ is given by (\ref{G4inG7}).

\section{Experimental tests}\label{exper}

In this section we consider two gravitational experiments in order to set bounds to the parameters 
of the model. First we look at a Cavendish type experiment. As a second test we compare the perturbed induced 
metric of the model with the generic PPN metric generated by a static  non rotating compact object.

\subsection{Cavendish type test}

In the context of the 5D Randall-Sundrum model, in \cite{Eingorn:2012yu} authors obtained the relative force 
corrections to the Newton's gravitational force between two massive spheres. The analysis was performed computing 
both the exact (considering the whole Kaluza-Klein massive tower) and the approximated gravitational potential 
(long distances limit) and comparing them in order to find out where the application of the approximate solution is 
appropriate. For their analysis they used the long distances limit of the potential generated by a massive particle (of 
mass $m$) in the RS model, which is of the form
\begin{equation}\label{ec.pote}
\varphi(r) \approx -\frac{mG_N}{r}\left(1+\frac{\alpha}{r^2}\right),\quad \alpha=l^2/2,
\end{equation} 
where  $l$ is proportional to the  anti de Sitter radius. This potential leads to the following gravitational force 
between two massive spheres
\begin{equation}\label{ec.fuerza}
F(r)=\frac{G_Nm_1m_2}{r^2}\left(1+\delta_F\right),
\end{equation} 
with
\begin{eqnarray}
\delta_F&=&-\frac{9\alpha}{8R^3R'^3}\nonumber\\
&&\left\{ \ln\left(\frac{r^2-\left(R'+R\right)^2}{r^2-\left(R'-R\right)^2}\right)
\left[-\frac{1}{4}r^4+\frac{1}{2}r^2\left(R'^2+R^2\right)-\frac{1}{4}\left(R'^2-R^2\right)^2\right]\right.\nonumber\\
 &&\left.  -r^2R'R+R'^3R+R'R^3 \right\} .
\end{eqnarray} 
Here $R$ and $R'$ are the radii of the spheres. Experimental data to verify this expression of the force 
is obtained from the Moscow Cavendish-type experiment \cite{Mitrofanov},  where one of the spheres was made  
of platinum with a radius $R\approx 0.087$ cm and mass  $m_1= 59.25\times 10^{-3}$ gr., whereas the second 
sphere was made of  tungsten with a radius  $R'\approx 0.206$ cm and mass $m_2= 706\times 10^{-3}$ gr. The  
center of the spheres were separated by a distance of  $r=0.3773$ cm.     

To obtain a bound on $l$, it is necessary to use an accurate value of Newton's gravitational constant. The values 
given by CODATA in 2010 \cite{Mohr:2012tt}  are
\begin{equation}
\frac{G_N}{10^{-11}}\frac{m^3}{kgs^2}=6.674215\pm 0.000092\quad\text{and}\quad 6.674252\pm 0.000124, 
\end{equation} 
here the relative error  $\Delta G_N/G_N$ shows the agreement of the measurements of the gravitational constant 
with the  $r^{-2}$  experiments \cite{Eingorn:2012yu}, i.e., the relation 
$\left|\Delta G_N/G_N\right|=\delta_F$ gives the upper limit for $\delta_F$, in order to not detect experimental 
deviations from the Newton's law. In the 5D RS model this implies that $l\leq 9.067\,\mu m$ and 
$l\leq 10.527\,\mu m$. A second approach using the complete solution gives   $l\leq 9.070\,\mu m$ and 
$l\leq 10.531\,\mu m$.  For practical use we can  take $l\leq 10\,\mu m$, which combined with the expressions 
(\ref{potential1}) and (\ref{potential2}) that we have obtained for the potentials in the brane produces a bound 
to the adS radius
\begin{eqnarray}
l^2&=& \frac{4\epsilon^2}{3}\Rightarrow \epsilon=\sqrt{\frac{3}{4}}l\simeq 0.86 l=8.6\,\mu m,\quad\text{for p=1},\\
l^2&=&\frac{5\epsilon^2}{6}\Rightarrow \epsilon=\sqrt{\frac{6}{5}}l\simeq 1.09 l=10.9\,\mu m,\quad\text{for p=2}.
\end{eqnarray}
These bounds are not in conflict with other previously reported in the literature, nevertheless the ones obtained here 
are weaker than for instance, the ones obtained by the Lamb shift, which gives bounds of the order 
$\epsilon \sim 10^{-14}m$ for $p=1$ and  $\epsilon \sim 10^{-13}m$ for $p=2$ \cite{MoralesTecotl:2006eh}.

\subsection{The four dimensional effective metric on the brane  }

Here we want to obtain the effective metric on the brane and look at the Newtonian and the parametrized post-
Newtonian (PPN) limits in order to set some bounds on the parameters of the theory. The PPN limit of metric theories 
of gravity contains 10 real valued parameters and to every metric theory of gravitation corresponds a set of values 
of the PPN parameters. The observational values of the parameters have been measured in the Solar System an 
also in binary neutron stars \cite{Will:1993ns,Will:2005va}.

The corresponding PPN metric in ``standard" spherical coordinates for a non rotating object  is (Will 2006)
\begin{equation}
ds_{PPN}^2= \left[  1- \frac{2G_Nm }{\rho} +\frac{ 2 G_N^2 m^2 (\beta- \gamma )}{ \rho^2}+ \ldots\right]dt^2 - 
\left[  1+ \frac{2G_Nm \gamma }{\rho}+ \ldots \right] d\rho^2 - \rho^2 d\Omega.
\end{equation}
For this case only the $\beta$  and  $\gamma$ parameters appear. The $\gamma$ parameter 
measures how much space curvature $g_{ij}$ is produced by unit rest mass, while   $\beta$  measures how much 
nonlinearity is there in the superposition law for gravity $g_{00}$. These two parameters are involved in the 
astrophysical effects of the perihelion shift and light deflection  as follows \cite{Hartle:2003yu}:
\begin{equation}
\delta_{prec} =  \frac{1}{3}(2 +2\gamma- \beta) \left[\frac{6 \pi G_N m}{c^2 a (1- e^2)}\right],
\end{equation}
here $a$ is the orbit's semi-major axis and $e$ is the eccentricity.
\begin{equation}
\delta_{def} =  \frac{1+ \gamma}{2} \frac{4G_N m}{c^2 b},
\end{equation}
in this case, $b$ is the impact parameter of the light ray.

The four dimensional effective metric on the brane  is given by
\begin{equation}
ds^2 = (1+ h_{00}) dt^2 + (-\delta_{ij }+ h_{ij}) dx^i dx^j, 
\end{equation}
For the cases  of one and two extra compact dimensions ($p$=1,2), taking into account the results (\ref{potential1}), 
(\ref{potential2}),  (\ref{potential3}) and  (\ref{potential4})  the metric is, to the lowest order that is needed here,
\begin{equation}
ds^2 =\left[ 1-\frac{2G_N m}{r}\left(1+ \frac{k_p}{r^2}\right)\right] dt^2 +
\left [-1+ \frac{2G_N m}{r} l_p \right]\delta_{ij }dx^i dx^j, 
\end{equation}
with
\begin{equation}
k_1= \frac{2 \epsilon^2}{3},\, \, \,  l_1=3; \, \, \, k_2=\frac{5 \epsilon^2}{12}, \, \, \, l_2=\frac{7}{3}.
\end{equation}
In spherical coordinates we have
\begin{equation}
ds^2 =\left[ 1-\frac{2G_N m}{r}\left(1+ \frac{k_p}{r^2}\right)\right] dt^2 +\left [-1- 2l_p \frac{G_N m}{r}\right]\left[ dr^2+ 
r^2 ( d\theta^2 + \sin^2(\theta) d\phi^2)\right].
\end{equation}
This metric is given in the isotropic form and we want it in the ``standard" form, which is the one adopted for the 
calculation of the PPN form of the metric of a static non rotating compact object. In order to obtain that form for our 
metric  we take the coordinate transformation 
\begin{equation}
\rho = r \left (1+ \frac{l_pG_N m }{r} +  \ldots\right ).
\end{equation}
The metric in the new coordinates is
\begin{equation}
ds^2= \left[  1- \frac{2G_Nm }{\rho} -\frac{ 2 l_p G_N^2 m^2 }{ \rho^2}+ \ldots\right]dt^2 - 
\left[ 1+ \frac{2 l_p G_Nm }{\rho} + \ldots \right] d\rho^2 - \rho^2 d\Omega.
\label{PPN1}
\end{equation}
The corresponding PPN metric is (Will 2006)
\begin{equation}
ds_{PPN}^2= \left[  1- \frac{2G_Nm }{\rho} +\frac{ 2 G_N^2 m^2 (\beta- \gamma )}{ \rho^2}+ \ldots\right]dt^2 - 
\left[  1+ \frac{2G_Nm \gamma }{\rho}+ \ldots \right] d\rho^2 - \rho^2 d\Omega.
\end{equation}
We notice by comparing the metrics that we do  have the Newtonian limit.  The values of the PPN coefficients 
$\beta $ and $\gamma $ for this theory are
\begin{equation}
\beta = 0; \, \, \,  \gamma =l_p.
\end{equation}
At the order of approximation considered here, the quantity $k_p$ does not appear, implying that the astrophysical 
tests do not impose a constraint on the anti de Sitter length.  The obtained values for the PPN parameters for this theory, 
in the cases where we have one or two extra compact dimensions, disagree with the observed values, since they are 
very close to one (the values for general relativity). We cannot tell if taking more compact dimensions will ameliorate  
the problem.

\section{Discussion}\label{conclu}

The perspective on known phenomena changes in light of models of spacetime that include extra dimensions. In 
particular,  brane world models  have provided new possibilities in high energy physics and cosmology to try to solve 
some problems like the hierarchy \cite{Randall:1999ee,Randall:1999vf} or dark matter/energy problems 
\cite{Maartens:2010ar,Maartens:2003tw}. However little
attention has been devoted to low energy physical effects which may shed light in regard to the viability of such 
higher dimensional scenarios by making reference to known experimental data including the Casimir effect, Lamb 
shift and others \cite{Frank:2008dt,Linares:2010uy,Linares:2011sg,MoralesTecotl:2006eh}. In fact even there some 
unexpected results may emerge as it is the case of non singular field configurations like the reported here and in 
previous works \cite{MoralesTecotl:2006eh,Linares:2011sg}.

In the present work we studied the gravitational potential produced by a source which looks pointlike to a 4D 
observer sitting in the single brane of an extended Randall-Sundrum-II scenario. Such source extends along the $p$ 
compact extra dimensions of the single brane thus forming a $T^p$ torus touching our usual 3D space at one point. A 
linear approximation for the hyper dimensional Einstein equations appropriate for such models was used.
We obtained a gravitational potential which is non singular at the position of the source in 4D. In line with our 
motivation we also calculated the gravitational force between two spheres in order to compare it with experimental 
data. This sets a bound for the adS radius of the order 10$\mu m$ which is consistent with previous more stringent 
electromagnetic results based on Lamb shift in hydrogen \cite{MoralesTecotl:2006eh}. On the 
other hand we obtained the PPN parameters for the field configuration corresponding to the {\em point like} source. 
The Newtonian limit is correctly contained in our results and this was proved explicitly for $p=1,2$ extra compact 
dimensions. However the PPN values obtained for the parameters of the RSII$p$ model  are out of range of the 
experimental data. This is not a problem as far as we do consider our brane model RSII$p$ as a test scenario rather 
than a realistic proposal to describe our world.

Future work along the lines we have developed here include the following. The gravitational radiation reaction 
problem may be reanalyzed in a setting similar to the one presented here. This may help to further understand the 
role of its specific features that allow to solve the divergent character of the standard 4D case. In particular it would be 
of interest to pin point what are the elements relevant for the resolution of the divergence in connection with the 
source, namely, whether is it linked to its topology, extension, codimension or something else. Further divergences in 
field theory may acquire a different form in brane worlds and we think they deserve some effort. This may be the case 
for instance for  quantum field theory in a brane world background.

\begin{acknowledgments}
RL, HAMT and LOP acknowledges partial support from PROMEP network: Gravitation and Mathematical Physics, 
under  grant UAMI-43, {\it Quantum aspects of gravity in cosmological models, phenomenology and geometry of 
space-time}. HAMT acknowledges partial support from grant CONACyT-NSF Strong backreaction 
effects in quantum cosmology. The work of LOP is part of the collaboration within the Instituto Avanzado de 
Cosmolog\'{\i}a.
\end{acknowledgments}

\bibliography{bibliography}
\end{document}